\documentclass[11pt]{article}

\usepackage{mathptmx}

\usepackage{amsmath,amssymb} 

\newtheorem{remark}{Remark}[section]

\newtheorem{lemma}{Lemma}[section]

\newtheorem{theorem}{Theorem}[section]

\newcommand{\proof}{\textsc{proof}\quad}
\newcommand{\qed}{\hfill \textsc{qed}}
            
\numberwithin{equation}{section}

\date{}

\begin{document}
\title{Mathematical quantum Yang-Mills theory revisited II: Mass without mass}
 
\author{Alexander  Dynin\\
\textit{\small   Professor  Emeritus of Mathematics, Ohio State University,}\\
\textit{\small  Columbus, OH 43210, USA}, \textit{\small  dynin@math.ohio-state.edu}}

\maketitle 
\begin{abstract}
Massless Dirac equation for spinor multiplets is minimally coupled with a unitary representation of an arbitrary
compact semisimple gauge group.  The  spectrum of the  quantized interaction Hamiltonian has a  positive mass gap running along the classical energy scale. 


\end{abstract}

\section{Introduction}
\subsection{Physics}
The concept of classically massless neutrinos is a cornerstone of the standard model of weak interactions.
However   an experimental evidence by T. Kajita  and A. McDonald groups (2015 Nobel prize) led
to a notion of oscillation between electronic and muonic neutrinos presumably because of   a difference in their masses. This is inconsistent 
 with the standard model \emph{unless} the masses have quantum origin.

F. Wilczek's \emph{QCD Lite}    is  the   Dirac -Yang-Mills dynamics (DYM)  of  the light $u$ and $d$  quarks   stripped of their   Lagrangian  mass terms.    Lattice simulation  of  interaction between  the massless quarks and massless gluons produces  (with no Higgs field) 99\% of  the  mass of visible  universe.

Quotation from Wilczek (see \cite[Subsection 1.2.1]{Wilczek}):
\begin{quotation}
My central points are most easily made with reference to two triplets and two anti-triplets 
of handed fermions, all with zero mass. Of course I have in mind that the gauge 
group represents color, and that one set of triplet and antitriplet will be identfied 
with the quark fields $u ,\ u_{\scriptscriptstyle R}$ and the other with $d,\ d_{\scriptscriptstyle  R}$. 
 
 Upon demanding renormalizability, this theory appears to contain precisely one 
parameter, the coupling [constant]. It is, in units  $\hbar = c = 1$, a pure number. I am ignoring 
the $\theta$ parameter, which has no physical content here, since it can be absorbed into 
the definition of the quark fields. Mass terms for the gluons are forbidden by gauge 
invariance. Mass terms for the quarks are forbidden by $SU(2)_{\scriptscriptstyle  L}\otimes SU(2)_{\scriptscriptstyle  R}$ flavor  symmetry.
\end{quotation} 

Besides the coupling constants $\kappa$, the classical  Lagrangians of quark flavors of the standard model  differ only   in their  phenomenological  quadratic   mass terms.  After  these mass terms are discarded, as in  the QCD Lite,  the Lagrangians  become indistinguishable.  

By \cite{Petermann}, physical constants of a quantum field theory, such as masses, are running  along  the classical energy scale. 

\subsection{Mathematics}
The present paper shows that the spectra of second quantized massless Yang-Mills-Dirac (DYM) Hamiltonians  (with arbitrary semi-simple compact gauge Lie groups) are discrete. In particular,
  the spectra have positive mass gaps. The sizes of the latter define the quantum masses. 

This paper is a   continuation of \cite{Dynin17} where
  the Clay Mathematics Institute "Quantum Yang-Mills theory" problem (see \cite{A}) has been solved in the case of  Yang-Mills
 fields in the vacuum. The problem requires 
    a mathematical proof  that
\begin{quotation}
 for any compact semisimple global gauge group, a nontrivial  quantum Yang-Mills theory exists on the four-dimensional Minkowski space-time and has a positive mass gap.
\end{quotation}

Accordingly,  the mass of  a classically  massless matter field is equal to the spectral mass gap of its quantum interaction with a gauge field associated with a global compact semisimple gauge Lie group.

As in \cite{Dynin17} the analysis is based on study of the Cauchy problem for coupled Dirac-Yang-Mills (DYM) equations. Because of the local  gauge symmetry, the Cauchy problem is  underdetermined: the solution set is covariant relative the action of the gauge  group. To reduce the gauge  degree of freedom one uses  a global gauge cross section, aka gauge (see \cite[Section I.1]{Faddeev}). Such is the Hamiltonian (aka temporal) gauge (see \cite{Segal}).
 
In this gauge the Cauchy problem for coupled Dirac-Yang-Mills (DYM) equations has a unique  global solution (Theorem \ref{pr:global} below).  

As in \cite{Dynin17}, the proof of existence involves  running  cutoffs of \emph{initial data} on the Euclidean balls $\|x\|<R$. This spatial renormalization converges as the radius $R$ increases to infinity.

   In the first order formalism the DYM equation becomes an infinite-dimensional Hamilton equation on the   manifold of solutions  of partial differential constraint equation \cite[Equation Chapter III, Equation (4.3)]{Faddeev}. 
 
 By R. Hamilton's tame implicit function theorem (see \cite{Hamilton}), the constraint  manifold carries  a global chart of a Frechet vector space (Theorem \ref{pr:constraints} below).

  The latter facilitates Berezin-Fock quantization cite{Berezin} of the DYM Hamiltonian with a positive spectral mass gap  (Theorem \ref{pr:spectrum} below).

\subsection{Nomenclature}
\label{pr:units}
To convert dimensional physical magnitudes into pure mathematical ones, 
the  natural units of quantum field theory are used:  Planck's  $\hbar$ (relevant for quantum effects), Einstein's $c$ (relevant  for relativistic effects), and
a  characteristic length unit, e.g.  fm (relevant at nucleonic  MeV energies). 

In this paper, \emph{smooth} means \emph{infinitely differentiable}.

The symbol * is used for a Hermitian conjugation. The \emph{bracketless} notation, say  $z^*w$, means
 $\langle z|w\rangle $.
 
 The Lorentz metric has the signature $(+,-,-,-)$.

\section{Classical DYM  Hamiltonian}
\subsection{Yang-Mills bosons}
The \emph{global gauge  group}  $\mathbf{G}$ of a  Yang-Mills theory is   a  
connected semisimple compact Lie group with the  Lie algebra $\mathbf{g}$ of skew-symmetric
 matrices   $X=-X^{\scriptstyle T }$. 

The Lie algebra carries the \emph{adjoint representation} $\mbox{Ad}(g)X=gXg^{-1}, g\in\mathbf{G}, X\in \mathbf{g}$, of the group $\mathbf{G}$ and the corresponding representation $\mbox{ad}(X)Y=[X,Y],\ X,Y\in\mathbf{g}$.  The adjoint  representation is orthogonal with respect to the \emph{positive  definite}  ad-invariant  scalar  product
\begin{equation}                                                                                            \label{eq:scalar}
X\cdot Y\ :=\ \mbox{trace}(\mbox{ad}X^T\mbox{ad}Y)\ =\ -\mbox{trace}(\mbox{ad}X\mbox{ad}Y), 
\end{equation}
the \emph{negative} of Killing form on $\mathbf{g}$.

 The \emph{local gauge  group} $\widetilde{\mathbf{G}}$ is the group of smooth $\mathbf{G}$-valued functions
 $\mathbf{g}(x)$ on $\mathbb{R}^{1+3}$ with the point-wise group multiplication.  The  \emph{local gauge Lie algebra}  
 $\tilde{\mathbf{g}}$ of $\mathbf{g}$-valued functions   on 
 $\mathbb{R}^{1+3}$   
consists of  infinitely differentiable $\mathbf{g}$-valued functions   on 
 $\mathbb{R}^{1+3}$ with the point-wise Lie bracket.   
 
$\widetilde{\mathbf{G}}$ acts via the pointwise adjoint action on 
the real vector space $\mathcal{A}$ of \emph{gauge   fields}   $A=A_\mu(x)\in\tilde{\mathbf{g}}$. 

 \smallskip  Gauge fields $A$ define   the \emph{covariant partial derivatives}  of  
\begin{equation}
\label{eq:covariant}
  \partial_{A,\mu}X\  :=
\  \partial_\mu X- \mbox{ad}(A_\mu)X,\quad
X\in\widetilde{A}.  
\end{equation}

Any $\tilde{g}\in\widetilde{\mathbf{G}}$ defines the affine \emph{gauge transformation}  
\begin{equation}                                                                                            \label{}
A_\mu\mapsto A_\mu^{\tilde{g}}:\ =\ \mbox{ad}\,(\tilde{g})A_\mu-(\partial_\mu \tilde{g})\tilde{g}^{-1},\ A\in \mathcal{A},
\end{equation}
so that $A^{\tilde{g}_1}A^{\tilde{g}_2}=A^{\tilde{g}_1\tilde{g}_2}$.

\medskip
Relativistic Yang-Mills \emph{curvature} $F(A)$ is  the 
antisymmetric tensor  
\begin{equation}
\label{eq:curvature}
F(A)_{\mu\nu} :=
\partial_\mu A_\nu-\partial_\nu A_\mu-[A_\mu,A_\nu].
\end{equation} 
The curvature is gauge invariant:
\begin{equation}
 \label{}
\mbox{Ad}\,(g)F(A)\ =\ F(A^{g}). 
 \end{equation}
 YM fields are solutions of the relativistic  Yang-Mills equation in the vacuum of the $\mathbf{g}$-valued  partial differential equation
 \begin{equation}
 \label{eq:YM}
 \partial_{A,\mu}F_{\mu,\nu}\ =\ 0.
 \end{equation}
 This is a relativistic  gauge invariant semi-linear  2nd order  
  partial differential $\mathbf{g}$-valued    equation  for one unknown $\mathbf{g}$-valued variable $A$.
   
  Every $\mathbf{G}$-orbit contains a connection $A$ with $A_0=0$ (see \cite{Segal}). 
  
   \emph{Henceforth to   reduce the gauge arbitrariness we impose this Hamiltonian (aka temporal) gauge condition}.

 The splitting $\mathbb{R}^{1+3}=\mathbb{R}\times\mathbb{R}^3$ implies $A_\mu=(A_0,A_j),\ j=1,2,3$, where $A_0$ are scalar fields and  $A_j$ are Euclidean  vector fields. This yields the splitting
\begin{equation}
\label{eq:splitting}
F^{\mu\nu}\ =\ \big(E^i:=F^{0j},\quad  B^i:=(1/2)\epsilon^{ijk}\partial_{A,j}\partial_{A,k}\big).
\end{equation}
The vector fields $E^i$ and $B^i$ are gauged versions of electric and magnetic fields. 
  
\subsection{Larks}
Let the  Minkowski  space-time $\mathbb{R}^{1+3}=\mathbb{R}\times\mathbb{R}^3$ be proper, i.e. time and space oriented.  Proper Lorentz transformations preserve both the space and time orientations; the orthochronos  Lorentz transformations preserve the time orientation but may flip the space orientation. Actually the 
orthochronous Lorentz group  is generated by proper Lorentz transformations and
 the improper Lorentz parity transformation of the spatial inversion.
\begin{equation}
P(x^0,x^1,x^2,x^3)\ :=\ (x^0,-x^1,-x^2,-x^3).
\end{equation}
The  group $\mathbf{SL}(2,\mathbb{C})$ of complex $2\otimes2$-matrices with the unit  determinant is the double cover group   of the group of proper Lorentz transformations. Two different $2\otimes2$-matrices cover the same proper Lorentz transformation if and only if they differ by sign.

Left-handed (vs. right-handed) Weyl spinors space $\mathbb{C}^2$ (vs. $\overline{\mathbb{C}}^2$)
carries the defining representations of the matrix group $\mathbf{SL}(2,\mathbb{C})$ and its  complex conjugate $\mathbf{SL}(2,\overline{\mathbb{C}})$. The corresponding left and right two-signed representations of the proper Lorentz group are irreducible, antidual,  and non-equivalent. They are intertwined by the representation of improper Lorentz parity transformation. 

\begin{remark}
Occasionally  the   group $\mathbf{SL}(2,\mathbb{C})$ is claimed to be isomorphic to the group $SU(2)_{\scriptscriptstyle  L}\otimes SU(2)^*_{\scriptscriptstyle  R}$.  However  the  Lie algebras  $\mathbf{sl}(2,\mathbb{C})$ and $\mathbf{su}(2)\times\mathbf{su}^*(2)$ of these groups  have different  real dimensions! 

Actually, $\mathbf{sl}(2,\mathbb{C})$ is the complex envelope   of  the real compact  semi-simple Lie algebra $\mathbf{su}(2)\times\mathbf{su}^*(2)$.  Thus, by the principle of analytic continuation (see \cite[Chapter VI, Section 41]{Zhelobenko}),    the tensor algebra of  finite-dimensional  representations of $\mathbf{su}(2)\times\mathbf{su}^*(2)$ is isomorphic via  analytic continuation   to the tensor algebra of their  holomorphic and  anti-holomorphic 
extensions to representations of  $\mathbf{sl}(2,\mathbb{C})$ .

The  representations algebra is generated by the fundamental representations of 
 $\mathbf{su}(2)\times \mathbf{su}^*(2)$. 
\end{remark} 
\emph{Massless Dirac fields} $\Psi=\Psi(t,x):\mathbb{R}^{1+3}\mapsto \mathbb{C}^2$ are  solutions of the \emph{massless} Dirac equation  
\begin{equation}
\label{eq:Dirac}
\gamma^\mu\partial_\mu\Psi=0,
\end{equation}
where $\gamma_\mu$ are Dirac $(4\times 4)$-matrices. The spinors $\Psi$ are subject to the Dirac conjugation $\overline{\Psi}:=\Psi^*\gamma_0$. \footnote{Actually the massless Dirac equation is the direct product of Weyl equations for left and right  2-spinor fields. However Weyl equations are not considered in this paper.}

Let the global gauge group $\mathbf{G}$ be an irreducible unitary representation of a compact semi-simple Lie group on  a finite multiplet  of massless Dirac spinor fields 
$\Psi\in \mathcal{W}^2(\mathbb{R}^{1+3},\mathbb{C}^2)$.

\emph{Lark fields} $\Lambda(t,x):=(\Psi,A)(t,x)$  are solutions of the  DYM partial differential  $\mathbf{g}$-valued equation consisting of minimally coupled  Dirac (\ref{eq:Dirac}) and Yang-Mills (\ref{eq:YM}) equations (see e.g. \cite[Section 2.]{Choquet})  \footnote{ "But O, Wreneagle Almighty, wouldn't un be a sky of a  lark" (see \cite[Page 383]{Joyce}).}
\begin{equation}
\label{eq:DYM}
 \gamma^\mu\partial_{A,\mu}\Psi=0, \quad \partial_{A,\nu} F_{\mu\nu}\  =\ J_{A,\mu},
\end{equation}
where  $J_{A,\mu}:=i\overline{\Psi}\gamma_\mu A\Psi$ is the source DYM current. The equation 
implies the gauged conservation law
\begin{equation}
\label{eq:conservation}
\partial_{A,\mu}J_{A,\mu}\ =\ 0.
\end{equation}
 \begin{remark}
 Larks with $\mathbf{G}= \mathbf{su}(2)\times\mathbf{su}^*(2)$  are gauge covariant massless  leptons.  Larks with  $\mathbf{G}$ = the fundamental representation of $\mathbf{SU}(3)$ are pairs of Wilczek  lite quarks.
\end{remark}
The DYM Equation (\ref{eq:DYM}) is equivalent to  the system of the first order partial differential $\mathbf{g}$-valued equations, the combination of  the system of dynamical \emph{evolution equations} 
\begin{eqnarray}
& &
\label{eq:evolution1}
\partial_t\Psi=i\gamma_0\gamma_k\partial_{A,k}\Psi,\quad \partial_t\overline{\Psi}\ =\ i\gamma_0\gamma_k\partial_{A,k}\overline{\Psi},\quad \partial_tA_k\ =\ E_k,
\\
& &
\label{eq:evolution2}
\quad\partial_tE_k\ =\ -(1/2)\epsilon_{klm}\partial_{A,l}B_m+J_k,\quad \partial_tB_k\ =\ (1/2)\epsilon_{klm}\partial_{ A,l}A_m,
\end{eqnarray}
and the non-dynamical system of \emph{constraint equations}
\begin{equation}
\label{eq:constraint}
B_k=(1/2)\epsilon_{klm}\partial_{A,l}A_m,\quad C(x):= \partial_kE_k-[A_k,E_k]+J_0\ =\ 0.
\end{equation}
The constraints are conserved on  solutions  of the evolution equations (\ref{eq:evolution1})
and (\ref{eq:evolution2})
in view of  Bianchi identity and Equation(\ref{eq:conservation}). 

The solutions set of Equation (\ref{eq:DYM}) is invariant under the action of the local gauge group 
$\tilde{\mathbf{g}}$. 
Every its orbit contains a smooth connection $A$ subject to \emph{Hamiltonian, aka temporal gauge} (see \cite{Segal}). 
Henceforth I  consider  only such connections. 
\begin{theorem}
\label{pr:global}
 There exists a unique smooth global solution  $\Lambda(t,x)$ of the Cauchy problem on Minkowski space-time for DYM equations with arbitrary smooth initial data 
\begin{equation}
\label{eq:data}
\lambda(x):=\big(\psi:=\Psi(0,x),\ a:=A(0,x),\   e:=E(0,x)\big).
\end{equation}

  A solution at $(t,x)$ is uniquely defined by their initial data $\lambda(x)$
 on the  spatial balls $\mathbf{B}_t:=\{x: |x|<t\}\subset\mathbb{R}^3$ (i.e. $B_t$ is the dependence domain of 
 $\Lambda(t,x)$ at  $(t,x)$).
 
  In particular  $\Lambda(t,x)$ are  pointwise limits of  solutions with   initial  data  cutoffs in   $\mathbf{B}_t$ as  $t\rightarrow\infty$. 
\end{theorem}
  \proof 
  The  Equations (\ref{eq:evolution1}) and (\ref{eq:evolution2}) is a first order hyperbolic semi-linear partial differential                                                                                                                                                                                                                                                                                                                                                                                                                                                                                                                                                                                                                                                                                                                                                                                                                                                                                                                                                                                                                                       
system for \emph{independent components} $(\Psi,B,A)$                                                                                                                                                                                                                                                                                                                                                                                                                                                                                                                                                                                                                                                                                                                                                                                                                                                                                                                                                                                                                                                                                                                                                                                                                                   with the  \emph{linear symmetric} principal part  and a polynomial  field $f$ as a non-linear term with $f(0)=0$. 
Let $\beta_r(t,x)\geq 0$ be a smooth cutoff function with compact support that is equal $1$ on  the solid cone $\{(t,x):|x|< r-t,\ 0< t< r\}$.
 
By \cite[Theorem 6.11 with  Note 3 and Theorem 7.3]{Mizohata}, the semi-linear system with cutoff nonlinearity 
$\beta f$ has a unique solution in the frustum 
\begin{equation}
\label{eq:frustum}
\mathbb{F}_r:=\{(t,x):|x|< r-t,\ 0< t\leq r/2\},
\end{equation} where $\beta_r f=f$. 

Furthermore,
the  one-to-one operator  fromsmooth initial data on the base $t=0$ of 
$\mathbb{F}_r$ to    smooth data on its top $t=r/2$  is a continuous map in the topology of the Frechet vector spaces. Then, by the open map theorem for Frechet spaces, the inverse operator  is continuous, i.e.  the Cauchy problems  on the frusta  are well-posed and their solutions
have the unit propagation speed.

Next, consider smooth Cauchy data $\lambda$ on the whole Euclidian space $\mathbb{R}^3$ with
no restrictions at the spatial infinity.

Let $\alpha_{2j}(x)\geq 0$ be  smooth functions with compact support in the balls $\mathbb{B}_{2j}=\{|x|<2j,\ j=1,2,...,\}$ such that $\alpha_j(x)=1$ on the balls $\mathbb{B}_{j}=\{|x|<j\}$. Then the solutions  of the Cauchy problem for the evolution system with the initial data $\alpha_j(x)\Lambda(0,x)$ have unique solutions $\Lambda_j(t,x)$ on the strips  $C_j:=\{(t,x):\  0<t<j,x\in\mathbb{R}^3\}$. 

As $j$ converges to infinity, the balls $\mathbb{B}_{j}$ increase to 
$\mathbb{R}^3$, the cylinders $C_j$ to the upper half $t>0$ of the Minkowski space, the initial datum  
$\alpha_j(x)\lambda(0,x)$  converges pointwise  to the initial datum $\lambda(0,x)$, and then 
$\lambda_j(t,x)$ converges to 
the unique global solution of the Cauchy problem for Equations  (\ref{eq:evolution1}),  (\ref{eq:evolution2})  on $\mathbb{R}^{1,3}$ constraint by Equation (\ref{eq:constraint}).

The system of DYM Equations    (\ref{eq:evolution1}),  (\ref{eq:evolution2}), and (\ref{eq:constraint})
  for fields $\Psi$,  $A$, $E$,  $B$  is equivalent to the shorter system  for $\Psi, A,E$
(see \cite[Equation 1.1]{Schwartz}, the combination of   dynamical \emph{evolution equations} 
\begin{eqnarray}
& &
\label{eq:evolution3}
\partial_t\Psi=i\gamma_0\gamma_k\partial_{A,k}\Psi,\quad \partial_t\overline{\Psi}\ =\ i\gamma_0\gamma_k\partial_{A,k}\overline{\Psi},
\\
& &
\label{eq:evolution4}
\partial_tA_k\ =\ E_k,\quad\partial_tE_k\ =\ -(1/2)\epsilon_{klm}\partial_{A,l}B_m+J_k,
\end{eqnarray}
and the non-dynamical system of \emph{constraint equations}
\begin{equation}
\label{eq:constraint1}
 C(x):= \partial_kE_k-[A_k,E_k]+J_0\ =\ 0.
\end{equation}
\qed

\begin{remark}
In the special case of YM equations (when $J=0$) Theorem \ref{pr:global} is due to
 \cite{Goganov} under the weaker condition that the initial datum belongs to \emph{local} Sobolev spaces $\mathcal{W}_2^3$. 
 For DYM equations a local version of Theorem \ref{pr:global} is due to \cite{Choquet} and  \cite{Schwartz}.  
 Actually, by \cite[Theorem 6.11 with Note 3 and Theorem 7.3]{Mizohata}, the global Theorem \ref{pr:global} holds under the same Sobolev condition.
 \end{remark}
The  short evolution system of (\ref{eq:evolution3}) and (\ref{eq:evolution4}) is   rewritten   as the    functional Hamiltonian equation  
  \cite[Chapter III, Equations (4.4)-(4.6)]{Faddeev} for fields
 \begin{equation}
\label{eq:fields}
\psi(t):=\Psi(t,x),\ a(t):= a_k(t,x),\ e_k(t):=F_{0k}(t,x),
\end{equation}
that are solutions of
\begin{eqnarray}
& &
 \label{eq:HDYM}
\partial_t\psi^* = i\partial_{\psi},\quad \partial_t\psi =- i\partial_{\psi^*}H,\quad 
 \partial_ta_k =\partial_eH,\quad \partial_te=-\partial_aH,\\
 & &
 \label{eq:Hamiltonian}
H:=\frac{1}{2}\int_{\mathbb{R}^3}\!d^3x\:\Big(i\psi^*\gamma_0\gamma_k\partial_{a,k}\psi+
\kappa^{-2}(\|\mbox{curl}^{(a)}a\|^2+\|e\|^2)),
 \end{eqnarray}
 where
 \begin{equation} 
 \mbox{div}^{(a)}e=\mbox{div}\,e-[a_k,e_k],\  (\mbox{curl}^{(a)}a)_k:=(\mbox{curl}\,a)_k-(1/2)\varepsilon_{ijk}[a_j,a_k],
\end{equation} 
and $\kappa$ is a dimensionless  \emph{coupling constant}. 
 
  The  time-independent  Hamiltonian functional $H$   is well defined by the initial values of (\ref{eq:fields})                                                                                   
 with compact supports  in the initial Euclidian space $\mathbb{R}^3$.  

Equation (\ref{eq:HDYM}) shows  that $(\psi^*=\psi^*(0,x), \psi=\psi(0,x))$ and
$(a=a(0,x), e=e(0,x))$ are  pairs of canonically conjugate variables with respect to the Poisson brackets on the Cauchy data defined via the Jacobi identity by the Poisson brackets 
\begin{eqnarray}
& &
\label{eq:Poisson}
\{H,\psi^*\}=\delta H/\delta \psi,\ \{H,\psi\}=-\delta H/\delta\psi^*,\\
& &
\{H,a\}=\delta H/\delta e, \{H,e\}=-\delta H/\delta a.
\end{eqnarray}

\subsection{DYM phase space} 
By equation (\ref{eq:covariant}), the dimensionality homogeneity implies  $a\propto L^{-1}$ in the natural units $c=1,\hbar=1$, where $L$ is a characteristic  length. Consequently, 
 \begin{equation}
 \label{eq:dimensionality}
a\propto L^{-1},\ e \propto L^{-2},\ \psi^*\ \propto L^{-3/2},\ \psi \propto L^{-3/2},\quad H\propto L^{-1}, \
C\ \propto L^{-2}.
 \end{equation} 
 Thus the $R$-scaling $x\mapsto Rx,\ 0<R<\infty,$ transforms
 the  Hamiltonian functional $H$ into $R^{-1}H$.

 By Theorem \ref{pr:global}, larks  $\Lambda(t,x)$  are uniquely defined by their Cauchy data 
 on the  spatial balls $\mathbb{B}(R):=\{x: |x|<R\}\subset\mathbb{R}^3$, and then, via the scaling 
 covariance, by the restricted  Hamiltonian functional on $\mathbb{B}:=\mathbb{B}(1)$
\begin{equation}                                                                                          
\label{eq:Hamiltonian1}
 H_{\mathbb{B}}:=\int_{\mathbb{B}}\!d^3x\:\big(i\psi^*
 \kappa^{-2}(\mbox{curl}^{(a)}a\|^2 +\|e\|^2)\big),
\end{equation}

 Sobolev-Hilbert spaces $\mathcal{A}^s$ of  connections $a(x)$ are the  completions of the spaces of smooth connections with compact supports in the open unit ball    
 $\mathbb{B}$  with respect to the norm squares 
\begin{equation}                                                                                            \label{eq:SH}
 \|a\|_s^2 :=
\int_{\mathbb{B}}\, dx\,\big(a(1-\triangle)^sa\big) < \infty. 
\end{equation}
The topological intersection $\mathcal{A}:=\cap\mathcal{A}^s$ is  a \emph{real}  Frechet  space, a subspace of the Hilbert space $\mathcal{A}^0$.

Similarly we define the real  Frechet  spaces  $\mathcal{E}:=\cap\mathcal{E}^s$ of connections $e(x)$,
$\mathcal{S}$ of real  scalar fields $u(x)$  on  $\mathbb{B}$ with values in $\mbox{ad}\:\mathbb{G}$,  the complex
Frechet vector space $\mathcal{D}$ of spinor fields on $\mathbb{B}$,  and their Hilbert space completions $\mathcal{E}^0(\mathbb{B}),\ \mathcal{S}^0(\mathbb{B}),\ \mathcal{D}^0(\mathbb{B})$.

By Sobolev imbedding theorem, the elements of $\mathcal{A},\ \mathcal{E},\ \mathcal{S}, 
\mathcal{D}$ are smooth fields.

By \cite[Corollaries 1.3.7. and 1.3.8.]{Hamilton}, 
$ \mathcal{A},\ \mathcal{E},\mathcal{S},\ \mathcal{D}$
are \emph{tame}    Frechet spaces.

Lastly, by \cite[Example 1.2.2.(3)]{Hamilton}, the semi-linear partial differential constraint self-transformation of 
$\mathcal{D}\times
\mathcal{A}\times\mathcal{E} $ 
\begin{equation}
\label{eq:mapH}
(\psi^*,\psi,a,e)\ \mapsto\  i\psi^*\gamma_0 a\psi\ +\ \mbox{div}\,e-[a_k,e_k]
\end{equation}
 is tame. Then the next  lemma is a corollary of  the R. Hamilton implicit function theorem \cite[Theorems 3.3.1 and 3.3.4.]{Hamilton}.
 \begin{lemma}
\label{pr:lemma}
If $(\psi^\diamond,a^\diamond,e^\diamond)$ is a solution of the constraint equation $c=0$ then there is 
a unique smooth tame map  $e(\psi,a)$
of a 
 neighborhood $\mathcal{N}^\diamond$ of $(\psi^\diamond,a^\diamond)$  
 in $\mathcal{D}\times
\mathcal{A}$ to $\mathcal{E}$ such that
\begin{equation}
\label{eq:nbhd}
C(\psi,a,e(\psi,a))\ =\ 0.
\end{equation}
\end{lemma}
\proof It suffices to check that the partial derivative with respect to $e$ of  the constraint functional 
$C(\psi^*,\psi,a,e)$, (which is linear in $e$)
\begin{equation}
\partial_eC\ =\ \mbox{div}^{[a]}e
\end{equation}
 is continuous with respect to $(\psi,a)$,  surjective, and has right inverse, 

Since Laplacian $\triangle$ commutes with $(1-\triangle)^s$, it is tame mapping from
$\mathcal{S}$ to itself. Since the inverse $\triangle^{-1}:\mathcal{S}^2\rightarrow 
\mathcal{S}$ exists, the operator  $\triangle$ is invertible in $\mathcal{S}$.

The gauge Laplacian $\triangle^{[a]} :=\ \|\mbox{div}^{[a]}\|^2$
  differs from the usual  Laplacian $\triangle$ by  a first order differential operator. Therefore it is a tame Fredholm operator  of zero  index in $\mathcal{S}$.

If $\triangle^{[a]} u=0$ then  
\begin{equation}
(\triangle^{[a]}u) u =  (\mbox{grad}^{[a]}\,u)(\mbox{grad}^{[a]}\,u)\ =\ 0,  
\end{equation}
so that  $\mbox{grad}\,u-[a,u]\ =\ 0$, or $\partial_ku\ =\ [a_k,u]$.

The computation
\begin{equation}                                                                                          
(1/2)\partial_k(u u)= (\partial_ku u)=[a_k,u] u= -\mbox{trace}(a_kuu-ua_ku)=0
\end{equation}
implies that  the solutions    $u\in\mathcal{S}$ are constant. Because they vanish on the ball boundary, they vanish on the whole ball.   Since the index of the Fredholm operator  $\triangle^{[a]}$ is zero,  its range  is a closed subspace with the codimension equal to the dimension  of its  null space. Thus    the operators 
\begin{equation}
\mbox{div}^{[a]}\mbox{grad}^{[a]}: \mathcal{S}\rightarrow \mathcal{S},\quad a\in\mathcal{A},
\end{equation} 
are surjective and
have the right inverse
$\mbox{grad}^{[a]}(\triangle^{[a]})^{-1}$.\qed

By their existence and uniqueness, the local maps $e=e(\psi,a)$  convert  the solutions set of the constraint equation (\ref{eq:constraint}) into a trivial bundle with the base $\mathcal{D}\times
\mathcal{A}$ and   fibers $\mathcal{E}_{(\psi,a)}$ that  consist of solutions for the equation
\begin{equation}
\mbox{div}\:e\ =\ -i\psi^*\gamma_0 a\psi\ +\ [a_k,e_k].
\end{equation}
The fiber $\mathcal{E}_\bot:=\mathcal{E}(\psi,0)$ consists of solutions for  the transversality equation $\mbox{div}\:e=0$.  
The term of  (\ref{eq:Hamiltonian1})
\begin{equation}                                                                                          
\label{eq:curvature}
\|\mbox{curl}^{[a])}a\|^2 +\|e\|^2\ = \ -F_{jk} F_{jk}
\end{equation}
 is the curvature of  the time-independent  gauge fields $a(x)$. Thus  $H$ is invariant under  smooth local  time-independent gauge group.

 By \cite[Proposition 1]{Dell'Antonio}, the closure of the local gauge Lie
group  $\widetilde{\mathbb{G}}^1$ in the Sobolev space $\mathcal{W}^1(\mathbb{B})$ is an infinite-dimensional \emph{compact group} with a continuous action in the Hilbert space  $\mathcal{A}^0$.
\footnote{$\widetilde{\mathbb{G}}^1$ is not a Lie group.} The  action orbits are compact so that 
the squared continuous Hilbert  norm $\|a\|^2$ has an  absolute minimal value on every orbit  which is  attained at a weakly transversal $a,\ \mbox{div}\:a=0$. 

By Sobolev embedding theorem, 
$\mathcal{W}^1(\mathbb{B})\subset\mathcal{W}^6(\mathbb{B})$. 
 Therefore the functional
$H$ has a unique continuation to  $\widetilde{\mathbb{G}}^1$-orbits in $\mathcal{D}^0\times \mathcal{A}^0\times \mathcal{E}^0$ and is \emph{constant} on each of them.
 
 Therefore the  space  of smooth constraint initial data has the total  Frechet vector space chart of the direct product 
 $\mathcal{E}_\bot\times(\mathcal{D}\times
\mathcal{A}_\bot)$ of weakly transversal fields with the flat parallel transport preserving the norm $\|e\|$.

All in all we have proved  
\begin{theorem}
\label{pr:constraints}
 DYM  Hamiltonian functional $H(\psi^*,\psi,a,e)$, constrained by Equation (\ref{eq:constraint}), is uniquely determined by its restriction to the Frechet  space $\mathcal{A}_\bot\times \mathcal{E}_\bot\times\mathcal{D}$.
\end{theorem}

Now modify  the DYM  Hamiltonian as the scaling invariant integral
\begin{equation}                                                                                            
\label{eq:N1}
H_R(\psi^*,\psi,a,e):=\  (R/2)\int_{\mathbb{B}(R)}\,d^3x\:(i\psi^*\gamma_0\partial_{a,k}\psi+\kappa^{-2}(\|\mbox{curl}^{(a)}\,a\|^2 + \|e\|^2 ).
\end{equation}
In view of  the scaling covariance let us set $R=1$.

Then the space of  \emph{dimensionless}  complex combinations of transversal Yang-Mills fields
\begin{equation}
\label{eq:dimension}
z\ :=\ (1/\sqrt{2})(a+ie),\quad  z^*\ := \ (1/\sqrt{2})(a-ie).
\end{equation}
admits the global   Frechet vector space chart $\mathcal{Z}:=\mathcal{A}_\bot\times\mathcal{E}_\bot\times\mathcal{D}^*_\bot\times\mathcal{D}_\bot$ on the phase space of the Hamiltonian system (\ref{eq:HDYM}).

\section{Quantum DYM Hamiltonian}

Fix  a  Yang-Mills field $a'\in \mathcal{A}_\bot, e'\in\mathcal{E}_\bot$ and consider the partial Hamiltonian
\begin{equation}
\label{eq:pm}
H\ =\  (1/2)\int_{\mathbb{B}}\,d^3x\:(i\psi^*\gamma_k\partial_{a',k}\psi+\kappa^{-2}(\|\mbox{curl}^{(a')}\,a'\|^2 + \|e'\|^2 )\  =:\ H'(\overline{\psi},\psi)+K,
\end{equation}
where $K$ is a constant.

 Replace $H$ with $H'$ since    the Hamiltonian dynamics on the  complex phase space $\mathcal{D_\bot}^*\times \mathcal{D_\bot}$ is the same if Hamiltonians   differ by   an additive constant.
 
 The Hamiltonian $H'$ is a Hermitian  real-valued quadratic form  of the  essentially selfadjoint 
 operator $i\gamma_k\partial_{a',k}$ on the domain $\mathcal{D}_\bot$.

Let $D$ denote its  selfadjoint closure  in $\mathcal{D}^0$. Since $D^2$ is a strongly elliptic second order partial differential operator on $\mathbb{B}$, the   spectrum of $D^2$, and therefore the spectrum of $D$,
 are   sequences  of finitely  multiple eigenvalues that converge to $\infty$ 
(see \cite[Section 1.3]{Ladyzhenskaya}).
 
Denote $\mathcal{F}$ the Fock functor (see \cite[Epigraph to Section X.7]{Reed}). Then 
the \emph{quantum Hamiltonian}  $\widehat{D}$  is an  unbounded operator in the fermionic Fock space 
$\mathcal{F}\mathcal{D}^0$ over the Hilbert space $\mathcal{D}^0$ 
By \cite[Section 6, Theorem 1]{Berezin}, the operator  $\widehat{D}$ is essentially self-adjoint in  $\mathcal{F}\mathcal{D}^0$ on the domain 
$\mathcal{F}\mathcal{D}_\bot$. 

\begin{theorem}
\label{pr:spectrum}
The spectrum of  $\widehat{H}'$ is sequence of finitely multiple eigenvalues converging to $\infty$, and therefore  has a mass gap, i.e. the zero spectral point is isolated.                                   
\end{theorem}
\proof
 $D$ is an orthogonal sum of the operators $D_+$, $D_-$, $D_0$ such that 
\begin{equation}
\label{eq:orthogonal}
i\overline{\psi} D_+\psi\geq 0,\quad i\overline{\psi} D_-\psi\leq 0,\quad i\overline{\psi} D_0\psi=0,\quad\ \psi\in 
\mathcal{D}.
 \end{equation}

Then the spectrum of $D_+$  is a sequence of positive eigenvalues of $D$ converging  to $+\infty$, 
the spectrum of $D_-$  is a sequence of negative eigenvalues of $D$ converging  to $-\infty$, and the spectrum of $D_0$ is the zero point (all eigenvalues are counted according to their multiplicity).

Let 
\begin{equation}
\label{eq:expansion}
\mathcal{D}^0\ =\ \mathcal{D}^0_+\oplus\mathcal{D}^0_-\oplus\mathcal{D}^0_0
\end{equation}
be the corresponding orthogonal expansion of the Hilbert space $\mathcal{D}^0$.

Since the functor $\mathcal{F}$ is  abelian  we have
\begin{equation}
\label{eq:F}
\widehat{D}\ =\ \mathcal{F}D_+\oplus \mathcal{F}D_0\oplus \mathcal{F}D_-.
\end{equation}
Let $\{e_k\}$ be an othonormal eigenbasis for $D_+$ with  eigenvalues $\lambda_k>0$; $\{e_l\}$ be an othonormal eigenbasis for $D_-$ with negative eigenvalues $\lambda_l<0$; and $\{e_m\}$ be an orthonormal basis for $D_0$ with the  eigenvalues $\lambda_m=0$. Note that $D_0$ has a finite rank.

 Then \begin{itemize}
\item   $\mathcal{F}D_+$ has an othonormal basis of
$n!^{-1/2}\wedge_{k=1}^{n}e_k,\ n\geq 0$, with the  eigenvalues  $\sum_{k=1}^{n}e\lambda_k,$;
\item
$\mathcal{F}D_-$ has an orthonormal eigenbasis of
$n!^{-1/2}\wedge_{l=1}^{n}e_l,\  n\geq 0$, with the  eigenvalues  $\sum_{l=1}^{n}\lambda_l$;
\item
$\mathcal{F}D_0$ has a finite orthonormal eigenbasis of
   $n!^{-1/2}\wedge_{m=1}^{n}e_m,\  n\geq 0$,
  with  zero eigenvalues. 
  \end{itemize}
 Thus spectrum of the quantum Hamiltonian $\widehat{H}'=\mathcal{F}D$ is a sequence of eigenvalues converging to $\infty$ and
 the vacuum  eigenvalue $0$ is an  isolated point of the spectrum of $\widehat{H}_{\mathbb{B}}$, so that the spectrum has  a positive mass gap.\qed

\begin{remark}
By the dilation covariance, the restricted Hamiltonian over the Euclidian ball $\mathbb{B}(R)$,
\begin{equation}
\label{Hamilton1}
H_{R}(\overline{\psi},\psi)\ :=\ \int_{\mathbb{B}(R)}\!d^3x\: i\overline{\psi} i\gamma_j\partial_{\kappa A,j}\psi\ =\ R^{-1}H_{\mathbb{B}}.
\end{equation}
Thus the mass gap of $\widehat{H}_{R}$ is running in the  inverse proportion to $R$, i.e. in the direct proportion to the effective  classical energy. Incidentally it depends on the selected $a'$.
\end{remark}


\begin{thebibliography}{20}
 \bibitem{A}
 \textit{Quantum Yang-Mills theory}, http://www.claymath.org/prizeproblems/index.html
 
 \bibitem{Berezin}
F. A.  Berezin, \textit{The method of second quantization} (Academic Press, 1966).

\bibitem{Dell'Antonio}
Dell'Antonio, G. and Zwanziger, D., \textit{Every gauge orbit passes inside the Gribov horizon},  Comm. Math. Phys., \textbf{138} (1991), 259-299. 



\bibitem{Dynin17}
 A. Dynin, \textit{Mathematical quantum Yang-Mills theory revisited}, Russian Journal of Mathematical Physics, \textbf{24}, (1), 26-43 (2017).
 
  \bibitem{Choquet}
Y. Choquet-Bruhat, D. Christodoulou, \textit{Existence of global solutions of the Yang-Mills, Higggs and spinor Weyl equations in 3+1 dimensions}, Annales scientifiques  de l'E.N.S., 4e serie, \textbf{14}, 481-506, 1981. 

\bibitem{Eardley}
D.M. Eardly,  V. Moncrief, \textit{The global existence of Yang-Mills-Higgs fields in 4-dimensional Minkowski space.II}, Commun. Math. Phys., \textbf{83}(1982),194-212.

 \bibitem{Faddeev}
L. Faddeev,  A. Slavnov,  \textit{Gauge fields, introduction to quantum theory}, addison-Wesley, 1991.

\bibitem{Goganov}
 Goganov , M. V., Kapitanskii,  L. V., \textit{Global Solvability of the Initial Problem for Yang-Mills-Higgs
Equations}, Zapiski LOMI \textbf{147}, (1985); J. Sov. Math. \textbf{37}(1987), 802-822.



\bibitem{Hamilton}
 R. Hamilton, \textit{The inverse function theorem of Nash-Moser}, Bulletin of the AMS, \textbf{7} (1982), 65-222.


 \bibitem{Joyce}
 Joyce,J., \textit{Finnegans Wake}, Wordsworth Classics, Various printings.
 

\bibitem{Ladyzhenskaya}
O. A. Ladyzhenskaya, \textit{The boundary value problems of mathematical physics} (Springer-Verlag, 1985).

  
 \bibitem{Mizohata}
 S. Mizohata, \textit{The theory  of partial differential equations}, Cambridge, 1973.
 
 \bibitem{Reed}
M. Reed, B. Simon,
\textit{II. Methods of modern mathematical physics}, Academic Press, 1972.
 
\bibitem{Schwartz}
G. Schwartz, J. Sniaticky,  \textit{Yang-Mills and Dirac fields in a bag, existence and
uniqueness theorems}. Comm. Math. Phys., \textbf{168}(1995), 441-453.

\bibitem{Segal}
I. Segal,  \textit{The Cauchy problem for Yang-Mills equations}, J. Func. Anal \textbf{33} (1979), 175-194.


\bibitem{Petermann}
 E. C. G. Stueckelberg,  A. Petermann,  \textit{La normalisation des constantes dans la theorie des quanta,}   Helv. Phys. Acta  \textbf{26}, 499-520 (1953).
 



\bibitem{Wilczek}
F. Wilczek,   \textit{Four Big Questions with Pretty Good Answers},  
Werner Heisenberg Centennial Symposium "Developments in Modern Physics" (Buschhorn, G.,  {W}ess,J., eds), 79-98, Springer, 2004.

\bibitem{Zhelobenko}
D. Zhelobenko,  \textit{Compact Lie groups and their representations}, AMS 1973.


\end{thebibliography}
\end{document}